\documentclass[aps,prl,twocolumn,nofootinbib,superscriptaddress,floatfix]{revtex4-2}

\usepackage[utf8]{inputenc}
\usepackage[T1]{fontenc}
\usepackage{amsmath,amssymb}
\usepackage{mathrsfs}
\usepackage{mathtools}
\usepackage{bm}
\usepackage{hyperref}
\hypersetup{colorlinks=true,citecolor=blue,linkcolor=blue,urlcolor=blue}

\newcommand{\Iplus}{\mathcal I^+}
\newcommand{\Q}{\mathcal Q}
\newcommand{\K}{\mathcal K}
\newcommand{\Dop}{\mathcal D}
\newcommand{\Ddual}{\widetilde{\mathcal D}}
\newcommand{\News}{\widehat N}

\newcommand{\dd}{\mathrm d}

\begin{document}

\title{\texorpdfstring{$w_{1+\infty}$}{w1+infinity} as the Frame Algebra
of Kerr Soft Dressing}

\author{Gabriel Menezes}
\affiliation{Departamento de Física Matemática - Instituto de Física, Universidade de São Paulo,\\R. do Matão, 1371, 05508-090 São Paulo -- SP, Brasil}
\email{gsm@if.usp.br}


\begin{abstract}
The Veneziano--Vilkovisky supertranslation is the residual large
diffeomorphism relating the canonical Bondi frame to the intrinsic frame
of the scattering bodies.  We show it leads a tower selected by the
exponentiating soft expansion, the object generating the Kerr multipoles
at three points.  Since \(e^{\eta\omega a\cdot q}\) splits into even and odd parts,
the tower alternates parity, and for aligned spin we solve it to all
orders in hyperbolic integrals.  After a chiral projection its
composition law is the classical $w_{1+\infty}$ bracket: the physical
content we assign to that algebra.
\end{abstract}

\maketitle

\paragraph*{Introduction and claim.}
Celestial symmetry arguments show that an infinite tower of soft graviton
modes realizes a \(w_{1+\infty}\) symmetry of four-dimensional
gravitational scattering at tree level, including processes with massive
particles \cite{Kapec:2016jld,Guevara:2021abz,Strominger:2021mtt,
Hamada:2018vrw,Li:2018gnc,Himwich:2023njb}.  In its original celestial form this algebra is most
directly seen as an organization of conformally soft insertions and their
operator products.  The conformal-primary basis and its soft-current
dictionary give the natural language for this perspective
\cite{Pasterski:2016qvg,Pasterski:2017kqt,Pasterski:2021rjz,
Donnay:2022aba,Hu:2023lso}.  Recent phase-space and Noether analyses
sharpen the question of how much of this foundation acts as a spacetime
symmetry of the radiative gravitational field
\cite{Freidel:2021ytz,Geiller:2024subleading,Donnay:2024carrollian}.  We
ask a more specific physical question: does the algebra control a directly
identifiable piece of black-hole scattering dynamics?  The answer proposed
here is yes, in a sharply delimited sector: the algebra acts on the
Kerr-selected soft dressings that turn the canonical Bondi frame into the
intrinsic frame appropriate to Kerr scattering.

This statement sits between several recent developments.  The amplitude
construction of Ref.~\cite{Elkhidir:2024ward} shows how BMS-frame
supertranslations are extracted from scattering amplitudes, while the
amplitudes program for gravitational-wave physics has turned on-shell data
into post-Minkowskian observables, radiation and waveforms
\cite{Cheung:2018wkq,Kosower:2018adc,Bern:2019nnu,Bern:2019crd,
Cristofoli:2021vyo,Herrmann:2021tct,Buonanno:2022pgc,
Aoude:2023vdk,Georgoudis:2023lgf,Georgoudis:2025memory}.  For spinning
compact objects this map is especially rigid, because the Kerr no-hair
relation packages the whole spin-multipole tower into
\(M_\ell+iS_\ell=M(ia)^\ell\), with \(a^\mu=S^\mu/m\) the specific-spin,
or ring-radius, vector, \(M=M_0=m\) the black-hole mass, and \(S^\mu\)
the spin vector.  If the soft dressing selected by the three-point
amplitude carries this same exponential, the associated asymptotic charge
is not merely a label on a soft theorem: it prepares the long-range Kerr
field, fixes the corresponding BMS frame shift, and selects observable
memory moments.  The generalized-BMS phase space makes smooth
\(\mathrm{Diff}(S^2)\) vector fields natural asymptotic data
\cite{Barnich:2010eb,Campiglia:2014yka,Campiglia:2015kxa,
Campiglia:2015qka,Compere:2018ylh}, higher-spin charges at null infinity
provide a covariant route to the same generators \cite{Freidel:2021ytz},
and the massive celestial analysis of Ref.~\cite{Himwich:2023njb} shows
that the \(w_{1+\infty}\) action is nontrivial even for massive hard
states.

One translation is important.  Globally the null-infinity construction
gives the real Poisson algebra of polynomial functions on
\(T^\ast S^2\), with the trace and curvature descendants needed to act on
shear data.  The celestial \(w_{1+\infty}\) algebra appears after a local
complex patch and a chiral polarization -- and that projection is where
our statement becomes literal, since the two parities recombine into a
single holomorphic tower.  This also fixes the relation to the infrared
triangle \cite{Strominger2014,He:2014laa,StromingerZhiboedov2014,
Strominger2017,Pasterski:2015tva,Choi:2024mgl} and to the
proper-observable analysis of Ref.~\cite{AndradeSpeziale:2026}, whose
smeared observables are the well-defined version of the memory probes
below.

\paragraph*{Veneziano--Vilkovisky frame dictionary.}
The starting point is not an abstract algebra but a concrete frame
ambiguity.  Veneziano and Vilkovisky showed that the intrinsic frame
naturally attached to the dynamics of particles and the canonical Bondi frame are
related by an $O(G)$ supertranslation
\cite{Bondi1962,Sachs1962,Veneziano:2022zwh}.  This
resolves the apparent mismatch between the order at which angular-momentum
loss appears in the intrinsic mechanical description
\cite{Damour:2020tta} and in the canonical asymptotic description.  For a
boosted linearized Schwarzschild source with momentum $p^\mu$, the
intrinsic-frame shear is
\begin{equation}
C^{\rm VV}_{AB}
=
-4G\,\frac{p^\mu p^\nu}{p\cdot q}\,
D_{\{A}q_\mu D_{B\}}q_\nu
\equiv -4G\,S^{(0)}_{AB}(p;q),
\label{eq:vv-shear-letter}
\end{equation}
where \(G\) is Newton's constant,
\(q^\mu(z)=(1,\hat{\boldsymbol q}(z))\) is the dimensionless null
direction associated with the unit spatial vector
\(\hat{\boldsymbol q}\), \(\gamma_{AB}\) is the unit-sphere metric, and
\(D_A\) its Levi--Civita derivative.  Curly sphere indices denote the
symmetric trace-free projection, for example
\(X_{\{A}Y_{B\}}=X_{(A}Y_{B)}-\frac12\gamma_{AB}X_CY^C\).
It is worth
being explicit that this is a residual gauge transformation, not an
abstract charge label.  After the small gauge fixing to Bondi gauge, the
transformations that survive at null infinity are generated by
\begin{equation}
\begin{split}
&\xi^u_T=T,\qquad \xi^r_T=\tfrac12D^2T+O(r^{-1}),\\
&\xi^A_T=-\tfrac1rD^AT+O(r^{-2}),
\end{split}
\label{eq:residual-vector-letter}
\end{equation}
and acting on \(g_{AB}=-r^2\gamma_{AB}-rC_{AB}+O(1)\) in our mostly-minus
signature these shift the shear by
\begin{equation}
\delta_T C_{AB}=-2\Dop_{AB}T,\qquad
\Dop_{AB}T=D_AD_BT-\frac12\gamma_{AB}D^2T.
\end{equation}
Everything below extends \eqref{eq:residual-vector-letter} to higher soft
orders.  At \(s=0\) the shift is an ordinary residual Bondi
diffeomorphism.  At \(s=1\) the charge is a parity projection of the
generalized-BMS Ward charge and becomes a spacetime diffeomorphism only
after the accompanying transformation of the celestial metric and the
homogeneous shear terms are restored.  For \(s\ge2\) we use canonical
transformations of the radiative phase space, without assuming that they
are induced by bulk vector fields.  We use the following sign convention.
The canonical frame is the Bondi
frame with \(C_{AB}|_{\Iplus_-}=0\), so the past limit of the Bondi
angular momentum agrees with the ADM value.  The
intrinsic frame is the center-of-mass scattering frame, where the early
shear is \eqref{eq:vv-shear-letter}.  We denote by \(T_{\rm VV}\) the
canonical-to-intrinsic parameter,
\(C^{\rm intrinsic}_{AB}=C^{\rm canonical}_{AB}-2\Dop_{AB}T_{\rm VV}\).
The inverse transformation, with parameter \(-T_{\rm VV}\), kills the VV
shear and returns to the canonical frame.  Therefore
\begin{equation}
\Dop_{AB}T_{\rm VV}=2G\,S^{(0)}_{AB}.
\label{eq:vv-frame-equation-letter}
\end{equation}
This equation is solved by a short sphere identity.  Let $x=p\cdot q$ and,
by Lorentz covariance, set $T=f(x)$.  Since $\Dop_{AB}x=0$,
\begin{equation}
\Dop_{AB}f(x)=f''(x)\!
\left(D_AxD_Bx-\frac12\gamma_{AB}D_CxD^Cx\right).
\end{equation}
The leading soft factor has the same trace-free tensor structure divided
by $x$, so \eqref{eq:vv-frame-equation-letter} gives
$f''(x)\propto 1/x$ and hence
\begin{equation}
T_{\rm VV}=2G\,(p\cdot q)\log(p\cdot q)
\label{eq:vv-solution-letter}
\end{equation}
up to ordinary translations.  Here and below \(\log x\) abbreviates
\(\log(x/\mu)\); changing the reference scale \(\mu\) adds a term
proportional to \(x=p\cdot q\), which is precisely such a translation.
The amplitude interpretation of
Ref.~\cite{Elkhidir:2024ward} then says that this same supertranslation is
the exponent of the leading gravitational Faddeev--Kulish dressing
\cite{FaddeevKulish1970,Ware:2013zja,Choi:2019rlz}.
The soft charge is the graviton displacement operator of that dressing;
it is constructed in the End Matter.  In the soft normalization used
below,
\begin{equation}
{\cal A}_{3,\eta}^{\rm Schw}(p+k\to p,k)
=\frac{\kappa}{2}\,(p\!\cdot\!\varepsilon_\eta)^2,
\qquad
S^{(0)}_\eta=\frac{{\cal A}_{3,\eta}^{\rm Schw}}{p\!\cdot q},
\end{equation}
where \(\kappa=\sqrt{32\pi G}\), \(\eta=\pm1\) is the graviton
helicity, \(k^\mu=\omega q^\mu\), and
\({\cal A}_{3,\eta}^{\rm Schw}\) is the scalar (Schwarzschild-source)
three-point graviton-emission amplitude.  The vector
\(\varepsilon_\eta^\mu(q)\) is the corresponding helicity polarization.
Sphere
kernels and helicity amplitudes are related throughout by the tangent
basis \(e^\mu_A=D_Aq^\mu\) and the electric spin-two projector
\begin{equation}
\begin{split}
E^{\mu\nu}_{AB}&=e^{(\mu}_Ae^{\nu)}_B
-\tfrac12\gamma_{AB}\gamma^{CD}e^\mu_Ce^\nu_D,\\
S^{(0)}_{AB}&=\frac{p_\mu p_\nu E^{\mu\nu}_{AB}}{p\cdot q},
\end{split}
\label{eq:projector-letter}
\end{equation}
so that \(S^{(0)}_{AB}=x^{-1}(D_AxD_Bx)^{\rm STF}\) with \(x=p\cdot q\).
The commutator of \(\Q_s\) with the linearised metric adds the boosted
Schwarzschild field, which the hard charge of
Ref.~\cite{Elkhidir:2024ward} converts into the shift
\(u\to u+T_{\rm VV}(q)\); this is the field-theory sense in which the
leading soft charge bridges the two frames.

\paragraph*{Soft charges, differential kernels and Kerr motivation.}
We now ask which part of the higher soft expansion extends this
dictionary.  With $C_{AB}$ the Bondi shear and $\News_{AB}$ the shifted
news \cite{Freidel:2021ytz,Ashtekar:1981bq}, the soft charge tower for a
rank-$s$ parameter $t^{A_1\cdots A_s}$ is
\begin{equation}
\Q^{\rm soft}_s(t)=
\int_{\Iplus}\dd u\,\dd^2z\,\sqrt\gamma\,
u^s\,\K^{(s)}_{AB}[t;C]\News^{AB}.
\label{eq:soft-charge-letter}
\end{equation}
Its field-independent kernel generates the inhomogeneous soft shift,
\begin{equation}
\K^{(s)}_{AB}[t;C]=\K^{(s,0)}_{AB}[t]+O(C),
\qquad
\nu^{(0)}_{t\,AB}=u^s\K^{(s,0)}_{AB}[t].
\end{equation}
The kernel parity is fixed, not chosen.  With an inverse-length
bookkeeper \(\lambda\) (set to the soft frequency \(\omega\) after the
level expansion), the classical helicity-\(\eta\) Kerr source is
\begin{align}
S_{\eta,{\rm exp}}(\lambda)
&:=S^{(0)}_\eta e^{\eta\lambda a\cdot q}
\nonumber\\
&=\sum_{s\ge0}\lambda^sS^{(s)}_{\eta,{\rm exp}},
\nonumber\\
S^{(s)}_{\eta,{\rm exp}}
&=S^{(0)}_\eta\frac{\eta^s(a\cdot q)^s}{s!}.
\label{eq:helicity-levels-letter}
\end{align}
so the two helicities differ only by the sign of the exponent, and
\begin{equation}
e^{\eta\lambda a\cdot q}
=\cosh(\lambda a\cdot q)+\eta\sinh(\lambda a\cdot q).
\label{eq:parity-alternation-letter}
\end{equation}
The reality factor is explicit: with a complex sphere dyad \(m_A\),
\(\gamma_{AB}=2m_{(A}\bar m_{B)}\), and
\(\mathcal T^{(s)}_{AB}=S^{(s)}_{+,{\rm exp}}m_Am_B+
S^{(s)}_{-,{\rm exp}}\bar m_A\bar m_B\), helicity conjugation gives
\(\overline{\mathcal T^{(s)}_{AB}}=(-1)^s\mathcal T^{(s)}_{AB}\).
We therefore match the real tensor
\begin{equation}
\mathscr S^{(s)}_{AB,{\rm exp}}:=
\begin{cases}
\mathcal T^{(s)}_{AB},&s\ {\rm even},\\
-i\,\mathcal T^{(s)}_{AB},&s\ {\rm odd},
\end{cases}
\label{eq:real-source-letter}
\end{equation}
the odd-line sign following the orientation convention fixed below.  These
are the electric and magnetic projections respectively, so the frame
equation fixes the electric source at even \(s\) and the magnetic at odd,
matching the Kerr no-hair relation (mass moments even in \(\ell\), current
moments odd).
For aligned spin the source lies entirely in the projected parity and
nothing is lost; for generic spin orientation a residual opposite-parity
piece survives at each level and is not fixed here.\footnote{Matching
\(\Dop_{AB}G(x,y)\) to \(\cosh(y)x^{-1}(D_AxD_Bx)^{\rm STF}\) forces
\(G_{xy}=G_{yy}=0\), hence \(G_{xx}\) independent of \(y\); this is
consistent only when \(y\) is a function of \(x\), i.e. for aligned spin.}
Writing the maximally longitudinal scalar of the relevant parity,
\begin{equation}
\chi_t^{(s)}:=
\begin{cases}
D_{A_1}\!\cdots D_{A_s}t^{A_1\cdots A_s}, & s\ \hbox{even},\\[2pt]
\epsilon^{BA_1}D_BD_{A_2}\!\cdots D_{A_s}t^{A_1\cdots A_s}, & s\ \hbox{odd},
\end{cases}
\end{equation}
so that \(\chi^{(0)}_T=T\) and \(\chi^{(1)}_Y=\epsilon^{AB}D_AY_B\),
the representative used in this Letter is
\begin{equation}
\K^{(s,0)}_{AB}[t]=
\begin{cases}
\Dop_{AB}\,\chi_t^{(s)}, & s\ \hbox{even},\\[2pt]
\Ddual_{AB}\,\chi_t^{(s)}, & s\ \hbox{odd},
\end{cases}
\label{eq:soft-kernel-differential-letter}
\end{equation}
where \(\Ddual_{AB}:=\epsilon_{C(A}\Dop_{B)}{}^{C}\), with trace and
curvature descendants belonging to the full global completion discussed in
a companion paper.  Thus \(s=0\) gives
\(\K^{(0,0)}_{AB}[T]=\Dop_{AB}T\), while \(s=1\) gives
\(\K^{(1,0)}_{AB}[V]=\Ddual_{AB}(\epsilon^{CD}D_CV_D)\).  The charge is
therefore an inverse problem: given the soft source, solve
\(\K^{(s,0)}_{AB}[t]=2G\,\mathscr S^{(s)}_{AB,{\rm exp}}\) for the corresponding
parity component of the generator \(t\).  The normalization is chosen so
that the \(s=0\) member is exactly
\eqref{eq:vv-frame-equation-letter}.
With the Ashtekar--Streubel bracket normalized accordingly,
\begin{equation}
\{C_{AB}(u,z),\Q_s^{\rm soft}(t)\}_{AS}
=u^s\K^{(s,0)}_{AB}[t]+O(C),
\label{eq:soft-inhom-action-letter}
\end{equation}
so the soft charge is the Hamiltonian generator of the inhomogeneous
frame shift.
The factor \(u^s\) is the moment conjugate to the soft frequency
expansion; with the regulator implicit in the zero-frequency limit,
\begin{equation}
\int\dd u\,u^s\News_{AB}(u,z)
=
i^s\left.
\partial_\omega^s
\big[\omega\,a_{AB}(\omega,z)+{\rm h.c.}\big]
\right|_{\omega=0},
\label{eq:soft-moment-letter}
\end{equation}
with \(a_{AB}\) the sphere-polarization projection of the graviton
annihilation operator.  Smearing with \(\K^{(s,0)}_{AB}\) inserts the
field-independent zero mode at the corresponding soft order.
The corresponding regulated memory probe
\(\mathfrak M_s[t;\rho]=\int\dd u\,\dd^2z\sqrt\gamma\,
\rho(u)u^s\K^{(s,0)}_{AB}[t]\News^{AB}\), with \(\rho\) specifying the
boundary prescription, is the observable measured by this
sector.
The identification proposed here is
\begin{equation}
\K^{(s,0)}_{AB}[t]
=
2G\,\mathscr S^{(s)}_{AB,{\rm exp}},
\label{eq:soft-kernel-identification-letter}
\end{equation}
where the right-hand side is the Kerr-selected exponentiating projection
of the universal soft contribution \cite{Weinberg1965,Cachazo2014,Bern:2014vva,
Campiglia:2016jdj,Laddha:2017ygw},
not the whole universal contribution and not the full soft theorem at that
order~\footnote{The form of
$\mathscr S^{(s)}_{AB,{\rm exp}}$ used here differs from that of
Ref.~\cite{Himwich:2023njb}: as explained in Ref.~\cite{Guevara:2018wpp},
subleading soft factors can be traded for the remaining powers so that the
series exponentiates, which is what we do.}.  This is not an arbitrary
prescription.  The soft charge is the Hamiltonian that displaces the
shear, so its kernel is the metric-frame shift, and equating it with the
exponentiating soft factor asks for the large-gauge generator whose soft
charge creates the Kerr-selected long-range field.  At \(s=0\) this
returns the VV supertranslation; at \(s=1\) the curl of a generalized BMS
vector; at higher \(s\) the representative of the parity fixed by
\eqref{eq:parity-alternation-letter}.  The same matching is required by
the Ward identity, since the hard charge below is represented on external
states by the same projected exponentiating operator.
At \(s=0\) the kernel is the usual supertranslation charge kernel and at
\(s=1\) its parity dual on the generalized-BMS vector; the higher-\(s\)
formula continues this pattern covariantly, tailored to the soft moments
whose metric multipoles are reconstructed from celestial data
\cite{Compere:2022cpm,Riva:2023xxm}.  It is not a special case of a
complete nonlinear BMS-flux formula; it is the Kerr-selected subsector of
the universal contribution whose hard representative is fixed by the
smeared soft theorem.
This restriction is physically motivated by Kerr black holes.  At the
amplitude level the spinning three-point object is the
Guevara--Ochirov--Vines exponential
\cite{ArkaniHamed:2017jhn,Guevara:2017csg,Guevara:2018wpp,Guevara:2019fsj},
whose classical limit is the source
\(S^{(0)}_\eta\exp[-i\omega W_\eta(q)]\) with
\(W_\eta=q_\mu\varepsilon_{\eta\nu}S_{\rm cl}^{\mu\nu}/(p\cdot\varepsilon_\eta)\)
and \(S_{\rm cl}^{\mu\nu}=\epsilon^{\mu\nu\rho\sigma}p_\rho a_\sigma\).
Helicity self-duality fixes the sign (End Matter): on three-point support
\(p\cdot k=0\),
\begin{equation}
e^{-i\omega W_\eta}=e^{\eta\omega\,a\cdot q},
\label{eq:W-self-dual-letter}
\end{equation}
exactly the classical shift in \eqref{eq:parity-alternation-letter}; this
same exponential is the Kerr multipole generating function
\(M_\ell+iS_\ell=M(ia)^\ell\)
\cite{Geroch1970,Hansen1974,Thorne1980}.  The conventions, the off-support
form of \(W_\eta\), and the multipole check are given in the End Matter.
\paragraph*{Leading and subleading checks.}
At $s=0$, \eqref{eq:soft-kernel-identification-letter} is exactly the VV
equation \eqref{eq:vv-frame-equation-letter}; the soft charge kernel
reproduces the leading soft factor and gives
\eqref{eq:vv-solution-letter}.  At $s=1$ the real source
\eqref{eq:real-source-letter} is magnetic, and the same Ward identity
fixes a sphere vector only through its curl,
\begin{equation}
\Ddual_{AB}\big(\epsilon^{CD}D_CV_D\big)
=2G\,\mathscr S^{(1)}_{AB,{\rm exp}}.
\label{eq:subleading-divergence-letter}
\end{equation}
This is already beyond standard or extended BMS: a generic solution of
\eqref{eq:subleading-divergence-letter} is not holomorphic on the sphere; it is a smooth generalized-BMS Ward parameter, a $\mathrm{Diff}(S^2)$
vector \cite{Campiglia:2014yka,Campiglia:2015kxa,Compere:2018ylh}, and
\eqref{eq:subleading-divergence-letter} is the magnetic projection of the
associated subleading soft Ward charge rather than the full residual
transformation of \(C_{AB}\) (see the companion paper).  Decomposing
\begin{equation}
V^A=D^A\Phi+\epsilon^{AB}D_B\Psi,
\label{eq:helmholtz-letter}
\end{equation}
we have $\epsilon^{AB}D_AV_B=D^2\Psi$.  The exponentiating sector therefore
fixes the magnetic potential $\Psi$ but not the electric potential $\Phi$,
consistent with the Kerr current dipole, the frame-dragging datum, being
odd-parity; even levels fix electric potentials and odd magnetic ones.  In
memory terms the leading displacement memory is fixed, and at subleading
order the magnetic projection of the first news moment, i.e. spin memory
\cite{Pasterski:2015tva}, with the centre-of-mass partner requiring
additional input; the all-orders treatment is in the companion paper.
The tower is not merely formally defined.  For spin aligned with the
spatial momentum it can be solved to all orders.  With
\(x=p\cdot q\) and \(y=a\cdot q\), both \(\ell\le1\) harmonics and hence
both annihilated by \(\Dop_{AB}\), the aligned configuration
\(p^\mu=(E,0,0,P)\) with \(m^2=E^2-P^2\), \(a\) the ring radius and
\(a\cdot p=0\), gives \(y=\alpha+\beta x\) where \(\beta=aE/mP\) and
\(\alpha=-am/P\).  The parity
split \eqref{eq:parity-alternation-letter} then reduces the whole tower to
two ordinary differential equations.  Writing
\(A=\lambda\alpha\), \(B=\lambda\beta\),
\[
\Phi_\lambda''=\frac{2G\cosh(A+Bx)}{x},
\qquad
\Psi_\lambda''=\frac{2G\sinh(A+Bx)}{x},
\]
their solutions modulo translations are
\begin{align}
\Phi_\lambda&=2G\big[\cosh A\,\mathcal C_B+\sinh A\,\mathcal S_B\big],
\nonumber\\
\Psi_\lambda&=2G\big[\sinh A\,\mathcal C_B+\cosh A\,\mathcal S_B\big],
\nonumber\\
\mathcal C_B&=\big[Bx\,\mathrm{Chi}(Bx)-\sinh(Bx)\big]/B,
\nonumber\\
\mathcal S_B&=\big[Bx\,\mathrm{Shi}(Bx)-\cosh(Bx)\big]/B,
\label{eq:aligned-solution-letter}
\end{align}
with \(\mathrm{Chi}\), \(\mathrm{Shi}\) the hyperbolic cosine and sine
integrals.  The electric mass-multipole tower sits in \(\mathrm{Chi}\) and
the magnetic current-multipole tower in \(\mathrm{Shi}\), and in this
configuration the projection is exhaustive.  The source used here is the
classical one, \(S^{(0)}_{AB}e^{\eta\lambda a\cdot q}\), justified by the
Newman--Janis complex shift and valid at generic \(q\); the amplitude form coincides with it for \(\lambda=\omega\) on three-point
support by \eqref{eq:W-self-dual-letter} and is its on-shell avatar.  Two checks: as
\(B\to0\),
\(\mathrm{Chi}(Bx)=\gamma_{\!E}+\log(Bx)+O(B^2)\), where
\(\gamma_{\!E}\) is the Euler--Mascheroni constant, so
\(\Phi\to2G\,x\log x\), recovering \eqref{eq:vv-solution-letter}; and
expanding in the spin, with \(w=\alpha+\beta x=a(Ex-m^2)/mP\) treated as a
single first-order quantity, gives
\(\Phi_\lambda=\sum_{r\ge0}\lambda^{2r}\chi^{(2r)}\) and
\(\Psi_\lambda=\sum_{r\ge0}\lambda^{2r+1}\chi^{(2r+1)}\), where the
level-\(s\) scalars
\(\chi^{(s)}\equiv\chi^{(s)}_{t_s}\) of the Kerr-selected parameters,
\begin{equation}
\begin{split}
\chi^{(s)}=\frac{2G\,a^s}{s!(mP)^s}\Big[&(-m^2)^s(x\log x-x)\\
&+\sum_{j=1}^{s}\binom{s}{j}
\frac{E^j(-m^2)^{s-j}x^{j+1}}{j(j+1)}\Big],
\end{split}
\label{eq:aligned-coefficients-letter}
\end{equation}
alternately electric and magnetic.  Every level carries a logarithm, with
residue \((-m^2)^s\), suppressed by \((m/E)^{2s}\) against the polynomial
terms; the pure-polynomial form \(2Gx^{s+1}/[s(s+1)s!]\) holds only in the
unphysical limit \(\alpha\to0\) at fixed \(\beta\).  Dimensionally,
\([\chi^{(s)}]=L^{s+1}\), where \(L\) denotes length, so each physical combination
\(\omega^s\chi^{(s)}\) has the length of a Bondi-frame shift; the
generating sums are not dimensionally homogeneous before the powers of
\(\lambda\) are included.

\begin{figure}[tb]
\includegraphics[width=\columnwidth]{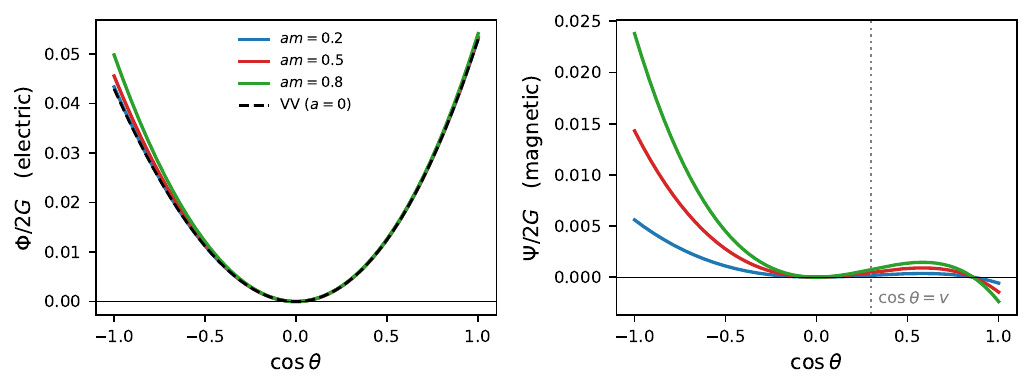}
\caption{Aligned-spin frame generators from
\eqref{eq:aligned-solution-letter}, normalised to vanish with their first
derivative at \(\cos\theta=0\) to fix the translation ambiguity, for
\(v=0.3\); the curves are labelled by \(\omega a\).  Left: the electric
generating potential \(\Phi_\omega\), collapsing onto the
Veneziano--Vilkovisky profile \(2G(x\log x-x)\) (dashed) as
\(\omega a\to0\).  Right: the magnetic generating potential
\(\Psi_\omega\), which exists only at nonzero spin.  The dotted line marks
the aberration angle \(\cos\theta=v\), where its source
\(\Psi_\omega''\), rather than \(\Psi_\omega\) itself, vanishes.}
\label{fig:tower}
\end{figure}

The tower therefore carries a concrete scale.  Since
\(\Phi''=2G\cosh(\omega w)/x\) and \(\Psi''=2G\sinh(\omega w)/x\), the
ratio of the spin-induced magnetic source at \(s=1\) to the
Veneziano--Vilkovisky electric source at \(s=0\) is exactly \(\omega w\),
and in the aligned configuration
\begin{equation}
\omega w=\chi\,(GM\omega)\,\gamma\,(v-\cos\theta),
\label{eq:spin-fraction-letter}
\end{equation}
with \(\chi=a/GM\) the dimensionless spin and \(v,\gamma\) the source
velocity and Lorentz factor.  Here \(\omega=2\pi f\) and \(M_\odot\)
denotes the solar mass.  For \(M=30\,M_\odot\), \(\chi=0.7\),
\(v=0.3\) and \(f=100\,\)Hz this reaches \(8.9\%\); for \(\chi=0.9\) and
\(v=0.5\) it reaches \(14.5\%\).  It vanishes at \(\cos\theta=v\), the
aberration angle marked in Fig.~\ref{fig:tower}, which is a sharp zero of
the magnetic source, and successive levels are suppressed as
\((\omega w)^s/s!\), so
the tower converges rapidly in the soft regime.

One may still recover the extended-BMS/celestial presentation by a local
chiral projection: complexify a patch of the sphere and keep only
\(V^z=Y(z)\), \(V^{\bar z}=0\), or more generally only holomorphic
symbols built from \(p_z\).\footnote{Here
\(p_z\) is the cotangent-fibre coordinate canonically conjugate to \(z\)
on \(T^\ast S^2\), not a spacetime momentum.}
This projection is useful for comparison with
celestial \(w_{1+\infty}\), but it is not the same as the real global Kerr
frame shift, whose vector is generically smooth and non-holomorphic.

\paragraph*{Hard charges and Ward identity.}
The charge acting on scattering states is
\begin{align}
\Q_s^{\rm total}(t)&=
\Q_s^{\rm soft}(t)+\Q_s^{\rm hard}(t),
\nonumber\\
\langle{\rm out}|[\Q_s^{\rm total}(t),S]|{\rm in}\rangle&=0.
\label{eq:ward-letter}
\end{align}
Here \(S\) is the scattering matrix.
The hard part is a flux/Ward operator: the null-infinity flux built from
the \(u^s\) moment of the hard stress tensor smeared with \(t\), whose
components and \(s=0\) reduction to the supertranslation phase
\cite{Elkhidir:2024ward} are given in the End Matter.

On external hard states the hard charge is represented by the hard soft
operator smeared with the same kernel \(\K^{(s,0)}[t]\); the Ward
identity is then the soft theorem in charge form, not an independent
all-order derivation of a new quantum symmetry.  The explicit representation, and the Kerr exponentiating piece of it, are given in the Supplemental Material below.

\paragraph*{Algebra after the charges.}
Only now do we ask what algebra these charges represent.  The linear soft
charges commute, since the field-independent kernels in
\eqref{eq:soft-charge-letter} have no functional derivative with respect
to $C_{AB}$, so \(\{\Q^{\rm soft}_{s,{\rm lin}}(t),
\Q^{\rm soft}_{s',{\rm lin}}(t')\}_{AS}=0\).
The non-abelian bracket enters through the homogeneous part of the
transformation; the derivation is given in the End Matter, and the
resulting parameter bracket is \eqref{eq:parameter-algebra-letter},
whose Hamiltonians close into \eqref{eq:charge-algebra-letter}.

The parity assignment is what makes this work at the lowest levels.  At
\(s=1\) the selected representative is divergence free, \(Y^A=\epsilon^{AB}
D_B\Psi\), and that subspace is preserved by the bracket,
\begin{equation}
\begin{split}
[Y_\Psi,Y_{\Psi'}]^A&=-\epsilon^{AB}D_B\{\Psi,\Psi'\},\\
\{\Psi,\Psi'\}&=\epsilon^{CD}D_C\Psi\,D_D\Psi',
\end{split}
\label{eq:sdiff-closure-letter}
\end{equation}
so the \(s=1\) Kerr sector is \(\mathrm{SDiff}(S^2)\), the group of
area-preserving sphere diffeomorphisms and the classical origin of
\(w_\infty\).  The electric representative would not close: the
bracket of two gradients is generically not a gradient.\footnote{In a flat
patch \(D(xy)\) and \(D(x^2/2)\) bracket to \((y,-x)\), which is
divergence free.}  Two limitations remain.  For \(s\ge2\) an STF rank-\(s\)
tensor carries two potentials while the inverse problem fixes one; and the
Kerr-selected points, parametrised by \((p^\mu,a^\mu)\), form a
finite-dimensional locus that \eqref{eq:sdiff-closure-letter} generically
leaves.  The algebra therefore acts on the space of frame data with Kerr
distinguishing a locus inside it.  This differs from the massive
deformation \(w_{1+\infty}(c)\), \(c\propto m\), found on the amplitude
side in Ref.~\cite{Guevara:2024color}: \eqref{eq:sdiff-closure-letter}
holds exactly at any hard mass.  A local chiral reduction of
\eqref{eq:parameter-algebra-letter} below, with
$F_{m,s}=z^{m+s-1}p_z^{s-1}$ and integer mode labels \(m,n\), gives
\begin{equation}
\{F_{m,s},F_{n,s'}\}
=
\big((s'-1)m-(s-1)n\big)F_{m+n,s+s'-2},
\end{equation}
the classical $w_{1+\infty}$ bracket up to spin-label conventions.  The
global celestial statement is more conservative: the algebra is the
polynomial Poisson algebra on $T^\ast S^2$, whose local or chiral
reductions are $w_{1+\infty}$-like
\cite{Kapec:2016jld,Guevara:2021abz,Strominger:2021mtt,
Himwich:2023njb,Freidel:2021ytz,Geiller:2024subleading,
Donnay:2024carrollian}.

\paragraph*{Back to the frame dictionary.}
Applying the logic of \eqref{eq:residual-vector-letter} to the linearized
Kerr--Schild field, the exponentiating three-point source produces the
intrinsic Coulombic Kerr representative to all orders in \(a^\mu\) at
leading order in \(G\), while the charges give the canonical/intrinsic
dictionary.  The claim is an asymptotic soft map, order by order in the
Kerr multipoles, not a full nonlinear all-PM transformation.  The leading
relation
\eqref{eq:vv-frame-equation-letter} implies
\begin{equation}
C^{\rm VV}_{AB}
=
-2\Dop_{AB}T_{\rm VV}
=
-4G\,S^{(0)}_{AB}.
\label{eq:vv-shear-from-potential-letter}
\end{equation}
Order by order in the soft frequency, the Kerr three-point factor replaces
the helicity source by
\(S^{(0)}_\eta\exp[-i\mathcal W_\eta(k)]
=\sum_{n\ge0}S^{(0)}_\eta[-i\omega W_\eta(q)]^n/n!\).
The real, parity-adapted recombination of its coefficients is
\(\mathscr S^{(s)}_{AB,{\rm exp}}\) in
\eqref{eq:real-source-letter}.
Combining \eqref{eq:vv-shear-from-potential-letter} and
the Kerr replacement above gives the Kerr-selected
canonical/intrinsic dictionary
\begin{align}
C_{AB}^{\rm intrinsic,Kerr}
&=
C_{AB}^{\rm canonical}
-2\sum_{s\ge0}\omega^s\K^{(s,0)}_{AB}[t_s],
\nonumber\\
\K^{(s,0)}_{AB}[t_s]
&=2G\,\mathscr S^{(s)}_{AB,{\rm exp}},
\label{eq:vv-kerr-dictionary-letter}
\end{align}
where the zeroth-order term is
\(T_{\rm VV}=2G(p\cdot q)\log(p\cdot q)\), with the kernel parity
alternating as in
\eqref{eq:soft-kernel-differential-letter}.  This is a generating
notation: at higher orders one must add the trace and curvature
completions rather than treat the dressed potential as an ordinary scalar
supertranslation.  Its higher powers in \(\omega W_\eta\) are the spin-multipole
frame shifts of the Kerr tower, electric at even order and magnetic at
odd.  The
\(w_{1+\infty}\)-like structure is therefore not a decorative symmetry
placed on top of this story; it is the classical symbol algebra acting on
these Kerr-selected soft dressings.
At \(s=0\) the inverse transformation \(-T_{\rm VV}\) kills the full VV
early shear.  At higher \(s\) the inverse flow kills only the
Kerr-selected component of the generalized intrinsic shear at
the parity fixed by \eqref{eq:parity-alternation-letter}.  It does not
claim to remove non-universal soft terms or nonlinear radiative shear.

The implication is concrete.  The celestial \(w_{1+\infty}\) tower has a
direct gravitational role: it acts on the soft frame shifts measured by
the regulated memory probe above, and after the chiral projection it is
their composition law.  The Kerr spin exponential is thus not only a
compact way of writing a three-point amplitude: at null infinity it
selects a parity-alternating tower of frame shifts, solved in closed form
by \eqref{eq:aligned-solution-letter} for aligned spin, whose observable
imprint begins with displacement memory at \(s=0\) and spin memory at
\(s=1\), followed by higher electric and magnetic memory moments.
Non-universal terms, radiative nonlinearities and possible
field-dependent extensions require a fuller phase-space analysis.

\paragraph*{Acknowledgments.}
We thank Leonardo Pipolo de Gioia for his collaboration at an early stage
of this project. We also thank Eduardo Casali, Radu Roiban, Fei Teng and Donal O'Connell for useful comments and suggestions. This work was partially supported by CNPq under grant
300767/2025-0 and FAPESP under grant 2025/02861-0.

\section*{End Matter}
\paragraph*{Kerr three-point amplitude and multipole check.}
This restriction is physically motivated by Kerr black holes.  At the
amplitude level the spinning three-point object is an operator on massive
spin states \cite{ArkaniHamed:2017jhn,Guevara:2017csg,Guevara:2018wpp,
Guevara:2019fsj}.  For a soft graviton \(k^\mu=\omega q^\mu\), the
Guevara--Ochirov--Vines convention is (\(\eta=\pm1\) is the graviton
helicity)
\begin{align}
\widehat{\cal A}_{3,\eta}^{(j)}
&=
{\cal A}_{3,\eta}^{(0)}
\exp\!\left(i\widehat{\mathcal W}_\eta\right),
\nonumber\\
\widehat{\mathcal W}_\eta
&=
\frac{k_\mu\varepsilon_{\eta\nu}J^{\mu\nu}}
{p\cdot\varepsilon_\eta}
=\omega\,\widehat W_\eta(q).
\label{eq:kerr-three-point-letter}
\end{align}
where \(j\) is the spin of the massive state,
\({\cal A}_{3,\eta}^{(0)}\) is the scalar minimal-coupling amplitude, and
\(J^{\mu\nu}=L^{\mu\nu}+S^{\mu\nu}\) is the total Lorentz generator on
the massive spinors.  Here \(L^{\mu\nu}\) is the orbital differential
generator and \(S^{\mu\nu}\) its intrinsic-spin part.  The physical
spin-\(j\) amplitude is the corresponding matrix element; the exponent
carries \(+i\widehat{\mathcal W}_\eta\), in
agreement with Ref.~\cite{Guevara:2018wpp}.  In the classical Kerr/source
limit, crossing to the outgoing-soft source convention used in the
dressing, this becomes
\begin{equation}
S_{\rm Kerr}^{\rm cl}(k)
=
S^{(0)}_\eta\exp[-i\omega W_\eta(q)],
\qquad
W_\eta(q)
=
\frac{q_\mu\varepsilon_{\eta\nu}S_{\rm cl}^{\mu\nu}}
{p\cdot\varepsilon_\eta},
\label{eq:kerr-source-soft-letter}
\end{equation}
so that \(\mathcal W_\eta(k):=\omega W_\eta(q)\) is dimensionless.  The
operator and source signs are therefore not being identified: at the
operator level one keeps
\(\exp(i\widehat{\mathcal W}_\eta)\) until the massive-spinor matrix
element is taken.  Expanding gives
\(\sum_{n\ge0}S^{(0)}_\eta[-i\omega W_\eta]^n/n!\).
For completeness, our spacetime conventions are
\(\eta_{\mu\nu}={\rm diag}(+,-,-,-)\),
\(\epsilon^{0123}=+1\),
\(S_{\rm cl}^{\mu\nu}=\epsilon^{\mu\nu\rho\sigma}p_\rho a_\sigma\),
\(\widetilde F^{\mu\nu}:=\frac12\epsilon^{\mu\nu\rho\sigma}F_{\rho\sigma}\),
and
\(\widetilde F_\eta^{\mu\nu}=-i\eta F_\eta^{\mu\nu}\) for
\(F_{\eta\mu\nu}=k_\mu\varepsilon_{\eta\nu}
-k_\nu\varepsilon_{\eta\mu}\).  They remove the apparent sign ambiguity:
\begin{equation}
\mathcal W_\eta(k)
=i\eta\left[
a\cdot k-\frac{(p\cdot k)(\varepsilon_\eta\cdot a)}
{p\cdot\varepsilon_\eta}\right].
\label{eq:W-self-dual-em}
\end{equation}
On three-point support \(p\cdot k=0\), hence
\(e^{-i\mathcal W_\eta}=e^{\eta a\cdot k}
=e^{\eta\omega a\cdot q}\), exactly the classical shift in
\eqref{eq:parity-alternation-letter}.
In the classical spin limit the same exponential is the multipole
generating function of linearized Kerr,
\begin{equation}
M_\ell+iS_\ell=M(ia)^\ell,
\label{eq:kerr-multipoles-letter}
\end{equation}
in the Geroch--Hansen/Thorne language
\cite{Geroch1970,Hansen1974,Thorne1980}. Here \(a\) in the scalar formula is
the signed magnitude of the ring-radius vector \(a^{\mu}\)  in the axisymmetric multipole frame.  Thus the part of the soft
expansion selected in \eqref{eq:soft-kernel-identification-letter} is
precisely the part known to package the Kerr multipoles.
It is also the natural higher-spin analogue of
\eqref{eq:soft-introduces-schwarzschild-letter}.  Replacing the scalar
three-point amplitude in \(Q_s\) by the spinning Kerr three-point
amplitude gives a coherent soft operator \(Q_s^{\rm Kerr}\) whose metric
commutator has the universal form
\begin{equation}
e^{-iQ_s^{\rm Kerr}}\hat h_{\mu\nu}(x)e^{iQ_s^{\rm Kerr}}
=\hat h_{\mu\nu}(x)+h_{\mu\nu}^{\rm Kerr,lin}(x),
\label{eq:soft-introduces-kerr-letter}
\end{equation}
after fixing the same asymptotic gauge and regulator prescription.  The
statement is not that a single fixed-\(s\) charge is the full Kerr field;
rather, the exponential above organizes the
linearized Kerr multipole tower term by term.  The check is explicit at
linear order: the dressing is linear in graviton operators, so the
Baker--Campbell--Hausdorff series truncates after the first commutator,
which is the Fourier transform of \({\cal A}_{3}^{\rm Kerr}\).  The
\(n=0\) term is boosted Schwarzschild, the \(n=1\) term the
frame-dragging current dipole, and higher powers reproduce the Kerr
multipole pattern.

\paragraph*{Charge bracket.}
The non-abelian bracket enters through the homogeneous part of the full
transformation,
\begin{equation}
\delta_t C_{AB}
=
\nu^{(0)}_{t\,AB}+\mathbb L_t^{(s)}C_{AB}+O(C^2).
\end{equation}
Here \(\mathbb L_t^{(s)}\) is the homogeneous linear differential operator
of the spin-\(s\) transformation: for \(s=1\) the usual tensor Lie
derivative up to trace and weight terms, and for higher \(s\) a
higher-derivative operator whose leading symbol is fixed by \(t\).  Its
existence for \(s\ge2\) is assumed rather than constructed; the bracket
below uses only the principal symbol and is insensitive to the remaining
terms.  The frame-dictionary results above do not depend on it.
The charge bracket follows from the commutator of the Hamiltonian flows,
not from the linear soft shifts alone: with
\(\delta_t\Phi=\{\Q_t,\Phi\}\) the Jacobi identity gives
\([\delta_t,\delta_{t'}]\Phi=\{\{\Q_t,\Q_{t'}\},\Phi\}\), and evaluating
on the shear at \(C=0\) leaves
\(\mathbb L_t^{(s)}\nu^{(0)}_{t'}-\mathbb L_{t'}^{(s')}\nu^{(0)}_t\).  The
fully symmetric top term cancels in this antisymmetrization, so the first
surviving universal term has degree \(s+s'-1\).
If \(\mathbb L_t^{(s)}=t^{A_1\cdots A_s}D_{A_1}\cdots D_{A_s}+\cdots\),
the principal symbol keeps the highest-derivative coefficient and replaces
\(D_A\) by a cotangent variable \(p_A\), so we associate to \(t\) the
polynomial \(F_t(x,p)=t^{A_1\cdots A_s}(x)p_{A_1}\cdots p_{A_s}\) on
$T^\ast S^2$.  Lower-derivative terms are responsible for trace,
curvature and ordering corrections, but the leading commutator is fixed
by this principal symbol.  The classical parameter bracket is
\begin{equation}
\{F_t,F_{t'}\}_{T^\ast S^2}=F_{[t,t']_\star},
\qquad
[\mathfrak g_s,\mathfrak g_{s'}]\subset
\mathfrak g_{s+s'-1}.
\label{eq:parameter-algebra-letter}
\end{equation}
Here \(\mathfrak g_s\) is the homogeneous degree-\(s\) subspace of
polynomial symbols.  On the full symbol space this is a genuine Lie
bracket, induced by the canonical Poisson bracket on \(T^\ast S^2\); STF
tensors are useful representatives but need not close by themselves
unless a trace-descendant prescription is specified.
Equivalently, denoting the principal-symbol map by \(\sigma\),
\(\sigma([\mathbb L_t^{(s)},\mathbb L_{t'}^{(s')}])
=\{F_t,F_{t'}\}_{T^\ast S^2}\).
Once the transformations close with parameter \([t,t']_\star\), their
Hamiltonians close into the charge with that parameter, up to a possible
boundary functional.
A companion paper gives the full component derivation, including the
lower-derivative completions and possible memory boundary terms; the
Letter keeps the universal principal-symbol part.
For $s=0,1$ this gives \([T,T']_\star=0\), \([V,T]_\star=V^AD_AT\) and
\([V,V']^A_\star=V^BD_BV'^A-V'^BD_BV^A\); for higher $s$ it is the
Schouten bracket of the leading symmetric symbols plus trace and curvature
descendants.

If the full transformations are Hamiltonian, the total charges obey
\begin{equation}
\{\Q_s^{\rm total}(t),\Q_{s'}^{\rm total}(t')\}
=
\Q_{s+s'-1}^{\rm total}([t,t']_\star)
+\mathcal K_{s,s'}[t,t';C],
\label{eq:charge-algebra-letter}
\end{equation}
where the possible extension is field dependent and tied to boundary shear
or memory data.  This is one of the central claims of the construction.

\paragraph*{Hard flux in components.}
The hard charge used above is the null-infinity flux
\begin{align}
\Q_s^{\rm hard}(t)&=\int_{\Iplus}\dd u\,\dd^2z\,\sqrt\gamma\,
\mathcal J_s[t]+\Q^{\rm hard}_{s,i^\pm}(t),
\nonumber\\
\mathcal J_s[t]&=u^s\,t^{A_1\cdots A_s}
D_{A_1}\!\cdots D_{A_s}T^{\rm hard}_{uu}
\nonumber\\
&\quad+s\,u^{s-1}t^{A_1\cdots A_s}
D_{A_1}\!\cdots D_{A_{s-1}}T^{\rm hard}_{uA_s},
\label{eq:hard-flux-em}
\end{align}
with \(\Q^{\rm hard}_{s,i^\pm}\) the massive-particle contribution entering
through the future/past timelike boundaries \(i^\pm\).
The quantities \(T^{\rm hard}_{uu}\) and \(T^{\rm hard}_{uA}\) are the
corresponding hard stress-tensor/flux components at null infinity.  At
\(s=0\) the density reduces to
\(T\,T^{\rm hard}_{uu}\), the standard supertranslation flux, and at
\(s=1\) to \(u\,Y^AD_AT^{\rm hard}_{uu}+Y^AT^{\rm hard}_{uA}\), the
generalized-BMS form; these are the only two levels at which an
independent check is available.  If finite-frequency gravitons are
included in \(T^{\rm hard}_{uu}\), the split between hard flux and
nonlinear radiative charge is conventional and must avoid double counting,
the total charge being the invariant object.  The \(u^s\) moment supplies
\(\partial_\omega^s\) at zero frequency while the angular derivatives act
on the hard momentum direction, so after integration by parts and LSZ
reduction one recovers the hard soft operator smeared with the same \(t\)
that appears in \eqref{eq:soft-kernel-differential-letter}.

\newpage

\vspace{1em}
\section*{Supplemental Material}
The following two constructions are standard and are not required to
follow the Letter's central argument; they are recorded here for
completeness.

\paragraph*{Soft charge and the Faddeev--Kulish dressing.}
More concretely, the low-frequency part of the three-point interaction
acts on a one-particle wavepacket as a displacement operator,
\begin{align}
|\psi\rangle&\longrightarrow e^{iQ_s}|\psi\rangle,
\nonumber\\
Q_s&\sim \sum_\eta\int d\Phi(k)\,
\delta(2p\!\cdot\! k)\,
{\cal A}_3(p+k\to p,k_\eta)
\nonumber\\
&\hspace{1.0cm}\times a^\dagger_\eta(k)+{\rm h.c.}
\label{eq:eor-soft-charge-letter}
\end{align}
Here \(d\Phi(k)\) is the Lorentz-invariant massless phase-space measure,
\(\eta=\pm\) labels graviton helicity, \(a^\dagger_\eta(k)\) creates a
soft graviton of momentum \(k^\mu=\omega q^\mu\), and
\({\cal A}_3\) is the scalar--scalar--graviton three-point amplitude.

The
commutator of \(Q_s\) with the linearized metric then adds the boosted
Schwarzschild field,
\begin{equation}
e^{-iQ_s}\hat h_{\mu\nu}(x)e^{iQ_s}
=\hat h_{\mu\nu}(x)+h_{\mu\nu}^{\rm Schw}(x),
\label{eq:soft-introduces-schwarzschild-letter}
\end{equation}
where \(\hat h_{\mu\nu}\) is the quantized linearized metric and
\(h_{\mu\nu}^{\rm Schw}\) the classical boosted linearized Schwarzschild
profile, up to the usual large-distance prescription.  The companion hard charge in
Ref.~\cite{Elkhidir:2024ward} then converts this Schwarzschild term into
the retarded-time shift \(u\to u+T_{\rm VV}(q)\).  This is the precise
field-theory sense in which the leading soft charge bridges the canonical
and intrinsic frames.

\paragraph*{Ward representation.}
On external hard states, the hard charge is represented by the
hard soft operator.  More explicitly, define
\(\mathcal D^{(s)}_{t,i}\) by smearing the leg-\(i\) hard soft operator
with the same kernel \(\K^{(s,0)}[t]\) that appears in the soft
charge.  After LSZ reduction,
\begin{equation}
\Q_s^{\rm hard}(t)|{\rm hard}\rangle
=
-\left(\sum_i \sigma_i\,\mathcal D^{(s)}_{t,i}\right)
|{\rm hard}\rangle ,
\label{eq:hard-state-representation-letter}
\end{equation}
where \(\sigma_i=\pm1\) distinguishes outgoing and incoming legs.  In the
same conventions, the unsmeared soft theorem is
\begin{equation}
\left.
\partial_\omega^s\big[\omega{\cal M}_{n+1}(\omega q)\big]
\right|_{\omega=0}
=
\left(\sum_i\sigma_i\,\mathcal D^{(s)}_{q,i}\right){\cal M}_n,
\label{eq:soft-theorem-derived-letter}
\end{equation}
Here \({\cal M}_n\) is the \(n\)-point hard amplitude,
\({\cal M}_{n+1}\) includes one additional outgoing soft graviton, and
\(\mathcal D^{(s)}_{q,i}\) is the corresponding unsmeared leg-\(i\)
soft differential operator.  Smearing with \(t\) gives
\eqref{eq:hard-state-representation-letter}.  The Ward identity used here
is therefore the soft theorem in charge form, not an independent
all-order derivation of a new quantum symmetry: it states
\(\langle{\rm out}|\Q_s^{\rm soft}(t)S|{\rm in}\rangle
=-\langle{\rm out}|\Q_s^{\rm hard}(t)S|{\rm in}\rangle\).
In the Kerr exponentiating sector the diagonal spin-dependent part of
\(\mathcal D^{(s)}_{t,i}\) is the smearing of
\(S_i^{(0)}(-iW_{\eta,i})^s/s!\) against \(t\), in the convention of
\eqref{eq:kerr-source-soft-letter}.

\end{document}